\documentclass[conference]{IEEEtran}
\usepackage[utf8]{inputenc}
\usepackage{url}
\usepackage{etoolbox}
\usepackage[skip=0pt]{caption}
\usepackage{fixltx2e}
\usepackage{amssymb}
\usepackage{amsmath}
\usepackage{graphicx}

\makeatletter
\patchcmd{\maketitle}{\@copyrightspace}{}{}{}
\makeatother
\begin{document}

\title{SignEPC : A Digital Signature Scheme for Efficient and Scalable Access Control in EPCglobal Network}

\author{\IEEEauthorblockN{Roy Ka-Wei LEE}
\IEEEauthorblockA{School of Information Systems\\
Singapore Management University\\
80 Stamford Road, Singapore 178902\\
Email: roylee.2013@smu.edu.sg}
\and
\IEEEauthorblockN{Yingjiu LI}
\IEEEauthorblockA{School of Information Systems\\
Singapore Management University\\
80 Stamford Road, Singapore 178902\\
Email: yjli@smu.edu.sg}}

\maketitle
\begin{abstract}
The EPCglobal network is a computer network which allows supply chain companies to search for their unknown partners globally and share information stored in product RFID tags with each other. Although there have been quite a number of recent research works done to improve the security of EPCglobal Network, the existing access control solutions are not efficient and scalable. For instance, when a user queries Electronic Product Code Information Service (EPCIS) for EPC event information, the EPCIS would have to query the Electronic Product Code Discovery Service (EPCDS) to check the access rights of the user. This implementation is not efficient and creates a bottleneck at EPCDS. In this paper, we design and propose a digital signature scheme, SignEPC, as a more efficient and scalable access control solution for EPCglobal network. Our paper will also evaluate SignEPC by considering the various possible attacks that could be done on our proposed model.

\end{abstract}





\section{Introduction}
A supply chain is a complex process which consists of a number of companies, such as suppliers, manufacturers, wholesales, distributors, retailers etc., working together to transform raw materials into products that can be consumed by consumers. Very often, the companies in a supply chain will have to collaborate and share information on their products so as to ensure that the right amount of products can be delivered to the consumers at the right places and in the right time. To achieve effectiveness and efficient in sharing of product information, EPCglobal network \cite{epcglobal} was proposed to allow companies in a supply chain to search for their unknown partners globally and share their product information.

Since the conception of EPCglobal network, researchers had proposed a number of privacy and access control features to improve the security of the network \cite{shi:platform}. However most of the existing access control solutions are inefficient and not scalable for real-world implementation. For example, in a simple Secure Electronic Product Code Discovery Service (EPCDS) model setup, the Electronic Product Code Information Service (EPCIS) will have to query EPCDS to check if the user has the access rights to the EPC event information requested. This is not efficient and scalable as it creates a bottleneck at EPCDS when the number of EPCIS and queries increases. In this paper, we design and propose a digital signature scheme, SignEPC, as a more efficient and scalable access control solution for EPCglobal network. We will also evaluate and demonstrate that SignEPC is able to prevent possible attacks effectively. The implementation and management of SignEPC will also be discussed in detail.

Our paper is organized as follows. Section~\ref{sec:background} provides background information on EPCglobal Network and some of the related works done to improve the security of the network. Section~\ref{sec:signepc} introduces SignEPC, the proposed digital signature scheme. Details on the implementation and management of SignEPC will also be discussed in this section. The analysis on how SignEPC prevents various possible attacks as well as the performance and scalability of SignEPC will be discussed in Section~\ref{sec:analysis}. Lastly, we conclude the paper in Section~\ref{sec:conclusion}.

\section{Background}
\label{sec:background}
In this session we briefly introduce the EPCglobal Network architecture and some of the common EPCDS models. Some of the existing access control solutions in the Secure EPCDS Model will also be discussed in this section.

\subsection {EPCglobal Network}
In EPCglobal Network, the products in a supply chain are attached with Radio Frequency Identification (RFID) tags, which contain EPC numbers that are globally identifiable. When a supply chain company receives products with RFID tags that contain EPC numbers, it stores the received products' EPC numbers in the repositories at company's website using EPCIS. Besides the EPC numbers, the company may also store EPC event information such as the time and location of the tag being read as well as other business related information. Companies may publish their service addresses or URLs to EPCDS. The EPCDS will act as a search engine and directory for supply chain companies to search for related partners as well as information about certain products \cite{kywe:evaluation}.    

\subsection {EPCDS Models}
\begin{figure}[h]
\begin{center}
\includegraphics[scale = 0.5]{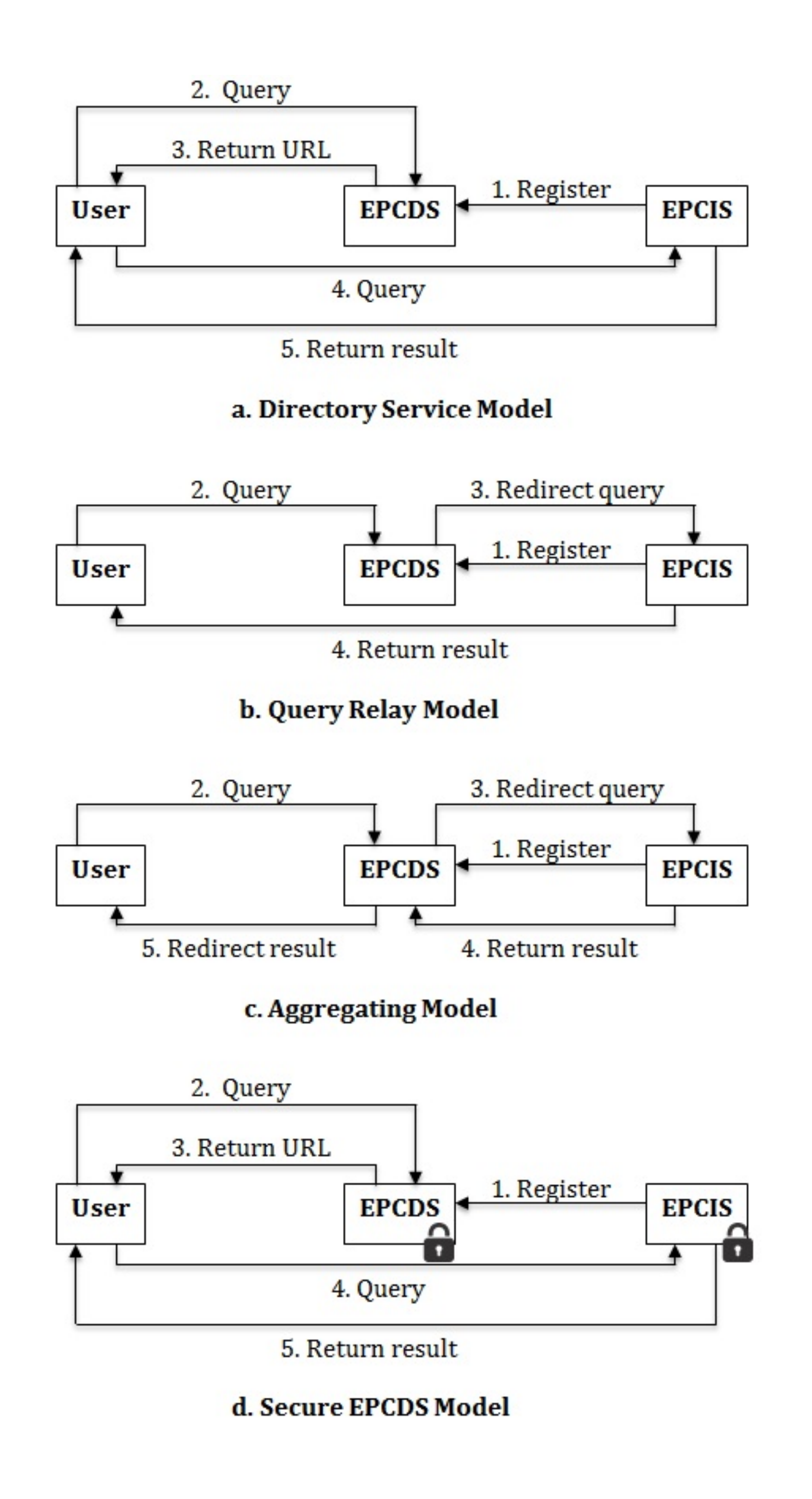}
\caption{EPCDS Models\label{fig:epcds}}
\end{center}
\end{figure}

A discovery service design or EPCDS model illustrates the interaction between the querying user, EPCDS and EPCIS \cite{kur:dis}. Figure~\ref{fig:epcds} shows the common EPCDS models, namely Directory Service Model, Query Relay model, aggregating model and most recently, Secure EPCDS Model which was proposed to enhance privacy and access control of information in EPCglobal Network.  

In the Directory Service Model, a user query the EPCDS with desired EPC number and the EPCDS returns the URLs of the related companies that handle product with the EPC number. The user then query the EPCIS of the related companies using the received URLs and the EPCIS return the EPC event information to the user. The returns of the URLs by EPCDS indicates ownership of the product information in that EPCIS companies. As such, every user knows exactly which EPC numbers are being handled by which EPCIS companies. This is a problem because many companies which consider their possession of items as confidential or sensitive information, do not want to expose their EPCIS URLs and were reluctant to publish them to EPCDS \cite{ben:sec}. 

In a Query Relay Model, the EPCDS does not return the URLs of the related companies. Instead, the EPCDS will redirect the queries from the user to the related companies and the EPCIS will return the relevant EPC event information to the user directly.

The Aggregating Model is similar to the query relay model. Instead of returning the relevant EPC event information directly to the querying user, EPCIS will return the result to EPCDS and the EPCDS will aggregate the result before returning it to the user. 

The Secure EPCDS Model is an enhanced version of the Directory Service Model. To tackle problem of leaking sensitive business information mentioned earlier in the Directory Service Model, the Secure EPCDS Model enforced access control policies in both EPCIS and EPCDS. For example, when a user query the EPCDS with a certain EPC number, the EPCDS will check the access rights of the querying user against the access control policies published by the EPCIS. If the querying user is authorized to access the EPC event information, the EPCDS will return the EPCIS URL to the user else it will return an error. Similarly, when a user query EPCIS using the received URL, EPCIS will query EPCDS to ensure that the querying user has the access rights to the EPC event information before returning results to the querying user. Although this simple access control mechanism is effective in ensuring that EPCIS URLs of companies as well as the EPC event information are not leaked to unauthorized parties, it creates a bottleneck at EPCDS as all access control queries are being channeled to it.

\subsection {Access Control in Secure EPCDS Model}
F. Benjamina and G. Oliver \cite{ben:sec} had highlighted many security challenges of EPCglobal Network. Since then many researchers have proposed different access control solutions to enhance the security of the network. The first simple access control policy in Secure EPCDS Model was proposed by J. Worapot et al. \cite{wor:des}. In that work, J. Worapot et al. suggested X.509 certificate as a authentication solution and three authorization rules: ``All'', which allow everyone to query and access the EPC event information,  ``Limited'', which only allows companies that own the same EPC number can access the EPC event information and hidden, and ``'Hide' which the event information is not shared with anybody. 

F. Kerschbaum \cite{ker:acc} proposed more complex access control policies with the Attribute Based Access Control (ABAC) model. He also suggested the visibility attribute for authorization, which is based on the partnership relationship between companies in the supply chain. There are basically three types of visibility attributes; ``Up-Stream'', where a company's EPC event information is made accessible to partners who published the same EPC number in EPCDS before the company,  ``Down-Stream'', where a company's EPC event information is made accessible to partners who published the same EPC number in EPCDS after the company, and ``Whole-Stream'', where the event information is made accessible to any partners who had published the same EPC number in EPCDS. J. Shi et al. \cite{shi:secds} had also extended the works and proposed adding subject attributes such as company ID and object attributes such as EPC numbers for authorization rules.

Using the above mentioned authentication and authorization rules, companies can define access control policies to better manage their EPC event information. For example, a manufacturing company may decide that a certain EPC number should only be made accessible to distributors and retailers by enforcing ``Down-Stream'' visibility attribute as its access control policies. Similar implementations can be done for up-stream and whole-stream partners in the supply chain. 

\section {Components in SignEPC}
\label{sec:signepc}
In this section we first introduce digital signature before describing the the implementation of our proposed solution, SignEPC. Implementation issues such as distribution of key and update of access policies will also be discussed and addressed in later part of this section. 

\subsection {Digital Signature}
A digital signature is a mathematical scheme allow users to ensures that an electronic document is authentic.  A valid digital signature have two properties:  authenticity, where the digital signature gives a recipient reason to believe that the message was genuinely created by a known sender, and non-repudiation, where the sender cannot deny that he or she have sent the message. 

The notion of a digital signature scheme was first conceived by W. Diffie and M. Hellman \cite{dh:cry}. R. Rivest et al. \cite{rsa:sign} subsequently invented the RSA algorithm, a public key cryptosystem which uses a pair of asymmetric keys. The RSA algorithm was also used to implement the first practical and usable digital signature scheme. Since then, digital signature was widely used in software distribution and financial sector to prevent forgery. 

The RSA signature scheme uses a pair of asymmetric keys; the private key, which is keep secret by the user and use to create a digital signature, and the public key which is distributed to other users to verify the digital signature. The asymmetric keys are related mathematically, however the parameters are chosen such that calculating the private key from the public key is computationally infeasible. The RSA public and private keys are also used in a way that satisfy the two digital signature scheme property; authenticity, when sender's private key is used to sign off a message so that the recipient believe that the message is genuinely from the sender, and non-repudiation when the sender is the only one with the private key and thus cannot deny that he or she have sent the message. The RSA signature scheme will be used in SignEPC.

\subsection {SignEPC Model}
\begin{figure}[h]
\begin{center}
\includegraphics[scale = 0.25]{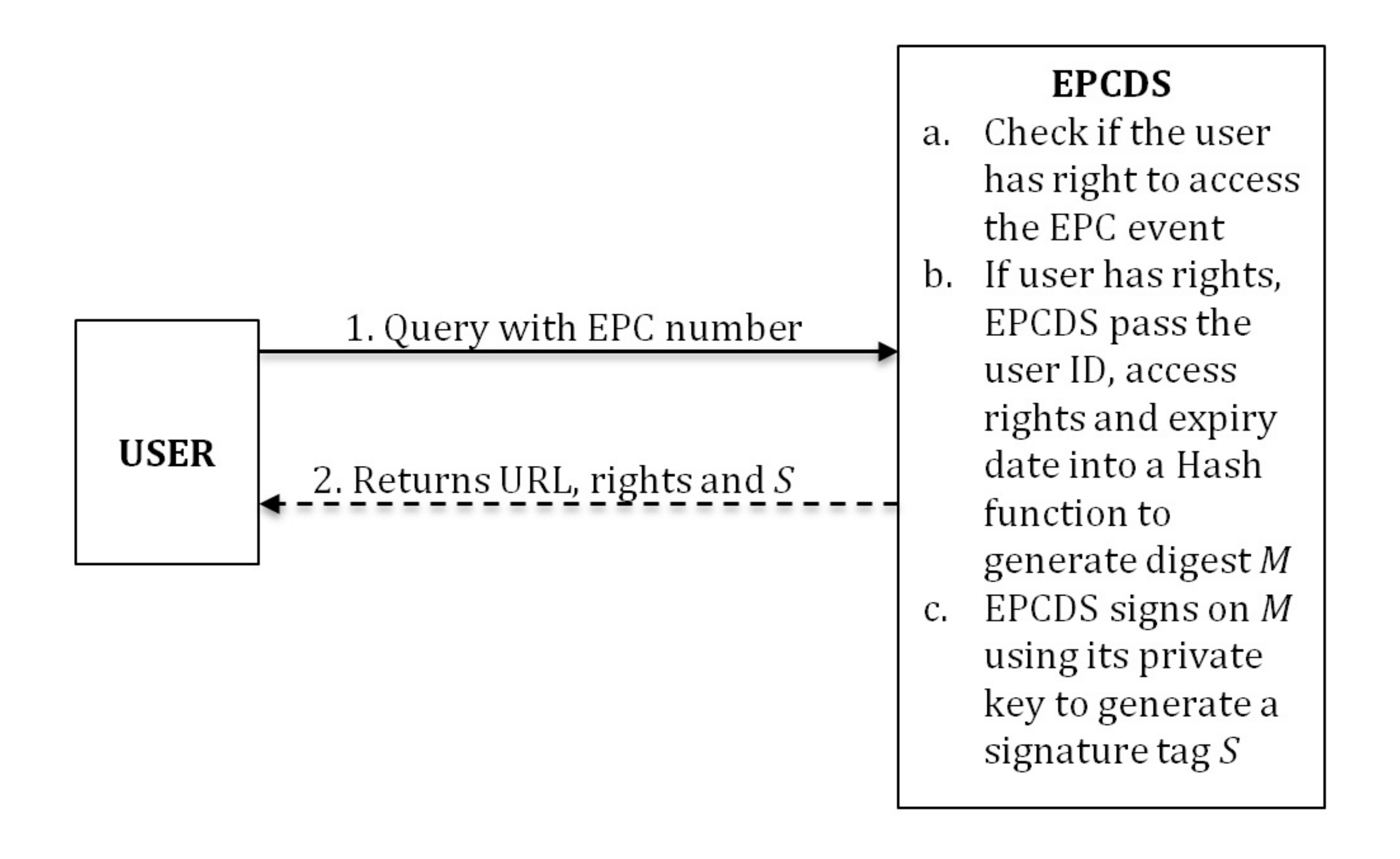}
\caption{EPCDS Sign on Access Rights \label{fig:sign}}
\end{center}
\end{figure}

In our proposed solution, SignEPC will leverage the properties of a digital signature scheme to solve the EPCDS bottleneck problem mentioned earlier. Figure~\ref{fig:sign} shows the process where EPCDS generates a signature tag \emph{S} using its RSA private key. The user first sends his query on a certain EPC number to EPCDS. The EPCDS checks if the user has the rights to access the EPC event information. Note that the access rights can be a simple concatenation of the EPC number and EPCIS url for user to access the full EPC event information or a more complex and fine grain access rights to specific EPC event information. For example, a EPCIS might grant a specific group of users to rights to access the warehouse where the product with the EPC number is stored but not the quality information. If the user has the access rights, EPCDS will pass the user ID, access rights and expiry into a hash function to generate a message digest \emph{M}, i.e. \emph{M} = \emph{Hash(userid,rights,expiry)}. The expiry is a datetime derived from the duration defined by EPCDS where a signature tag will remain valid. This duration is also being made known to EPCIS for signature verification. EPCDS then signs on the message digest using its private key \emph{d} to generate a signature tag \emph{S}, i.e. \emph{S}=\emph{sign\textsubscript{d}(M)}. Finally, the EPCDS returns the URL, user's access rights, signature tag \emph{S} to the querying user.

\begin{figure}[h]
\begin{center}
\includegraphics[scale = 0.25]{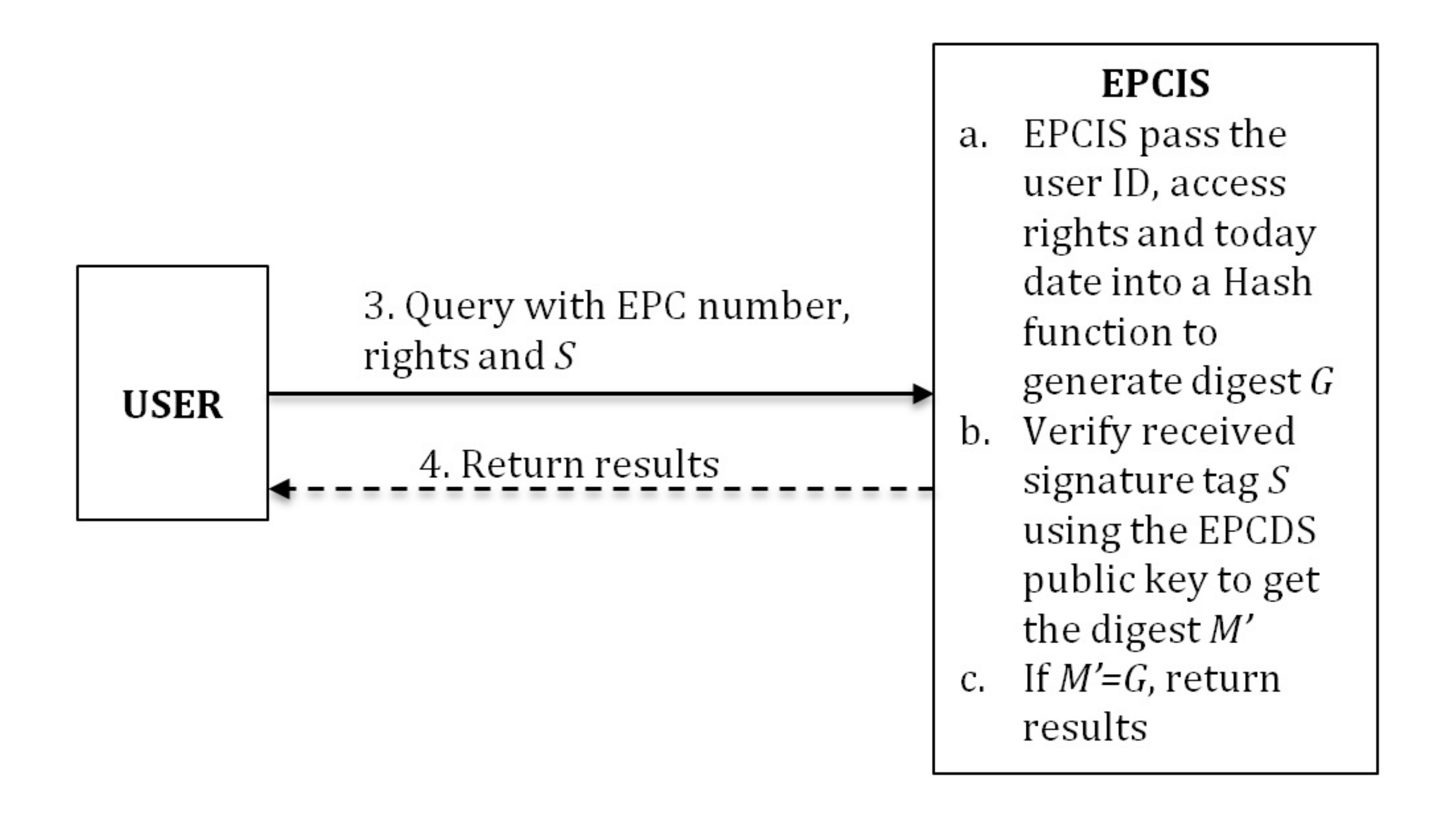}
\caption{EPCIS Verify Access Rights \label{fig:verify}}
\end{center}
\end{figure}

Figure~\ref{fig:verify} shows the signature verification process performed by EPCIS. The EPCIS first receives the query, user access rights and signature tag \emph{S} from the querying user. Assuming that the EPCDS defined expiry duration of a signature is set to daily, i.e. the validity of a signature tag is till 2359H of the day it is issued, the EPCIS will pass the querying user ID, received access rights and today's date into a hash function to generate message digest \emph{G}, i.e. \emph{G} = \emph{Hash(userid,rights,todaydate)}. EPCIS then verifies the signature tag \emph{S} by generating a message digest \emph{M'} using the EPCDS public key \emph{e}, i.e. \emph{M'} = \emph{verify\textsubscript{e}(S)}.  If \emph{M'} = \emph{G}, EPCIS can be sure that the message digest sent from the querying user is not being tampered with and EPCIS verifies that the user has the access right to the EPC event information. Finally, the EPCIS return the EPC event information to the querying user.

With the above described mechanism, the SignEPC model allows EPCIS to verify the access rights of the querying user without having to query the EPCDS, thus solving the access right bottleneck problem.

\subsection {Distribution of Keys}
\begin{figure}[h]
\begin{center}
\includegraphics[scale = 0.3]{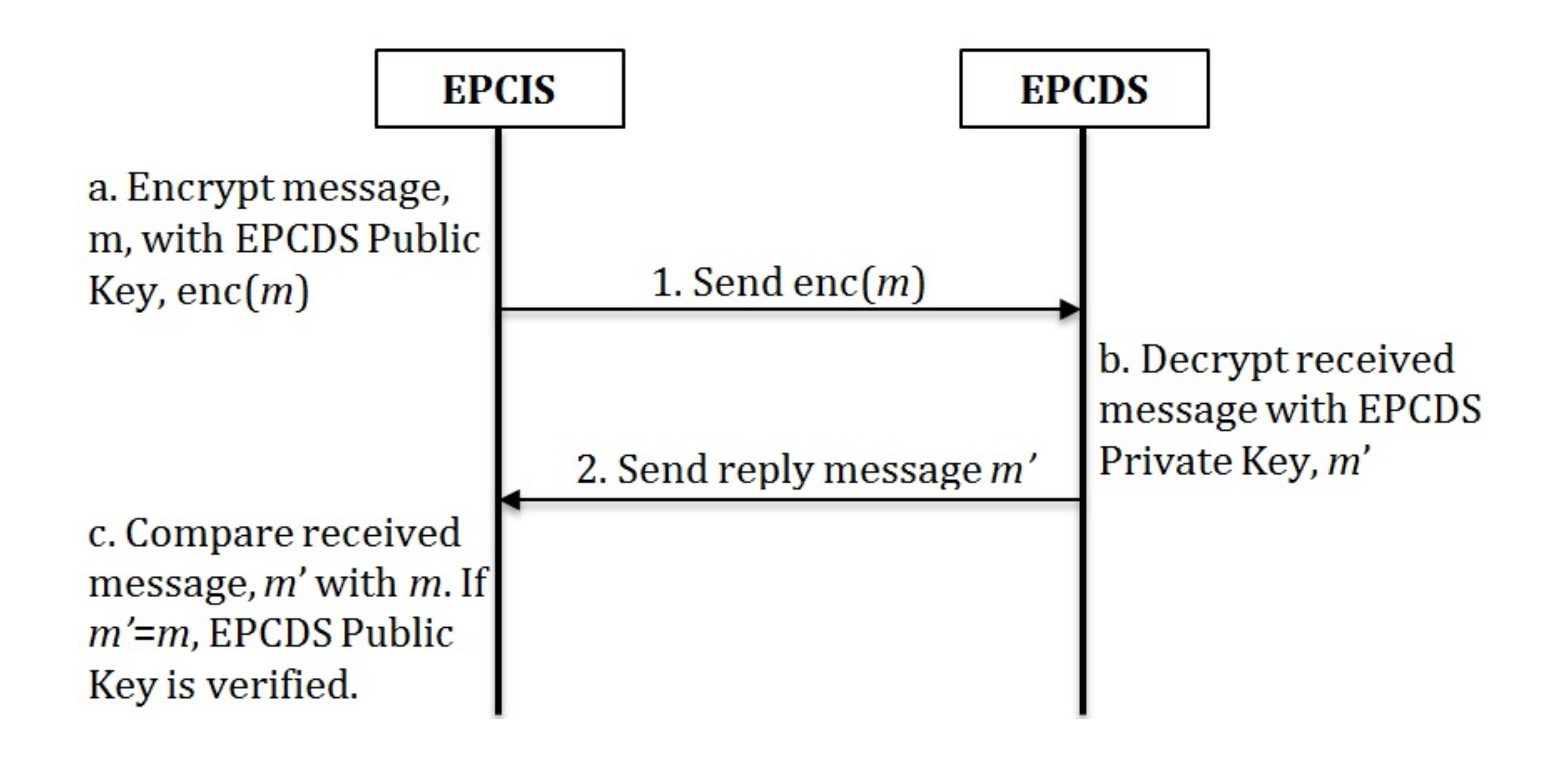}
\caption{Verify EPCDS Public Key \label{fig:key}}
\end{center}
\end{figure}

The Public Key Infrastructure (PKI), which is an arrangement that binds public key to an unique user by the means of a certificate authority (CA), is commonly used in the distribution of public keys over the web \cite{rfc:2560}. In the context of SignEPC, we assume that the EPCDS is an trusted entity and it can assume the role of CA. The EPCDS can publish and make available its public key to all users and the EPCIS can verify if a public key belong to EPCDS by issuing a challenge message to the EPCDS. Refer to Figure~\ref{fig:key} for the EPCDS public key verification.

The EPCIS may also schedule routine checks on the expiry of the current EPCDS public key and renew it accordingly. With the EPCDS public key verified, EPCIS can leverage on the SignEPC model to verify the access rights of the querying users.

\subsection {Updates on Access Policies}
The SignEPC model is also able to handle changes in access policies efficiently. Consider the scenario where the EPCIS publishes a change in its access policies for a certain EPC number to EPCDS, and the new access policies invalidate the access for a user. The user with expired access rights may attempt to query the EPCIS using the past EPCDS issued tag. 

This scenario is however prevented by expiry date feature of SignEPC; if the user attempt to use past EPCDS issued tag and the date of query is after the expiry date of the issued tag, EPCIS will reject the use query and not return any results. An assumption made in this scenario is that the generated tag has expired when the user made the query, if the generate tag has not expired and the access rights for the user has changed, the user could still succeed in querying the EPCIS using the past EPCDS issue tag. It is thus important to keep the expiry window of the EPCDS issue tag small or time the update on access policies close to the signature tag expiry date so that most of the past EPCDS issued tags would be expired as soon as possible. 

\section {Analysis of SignEPC}
\label{sec:analysis}
In this section, we analyse SignEPC in a two aspects. Firstly, we analyze whether the solution will be able to prevent possible attacks that could be done on SignEPC. Next we evaluate the performance and scalability of our solution in comparison to the the Secure EPCDS model design. 

\subsection {Attacking SignEPC model}
\begin{figure}[h]
\begin{center}
\includegraphics[scale = 0.23]{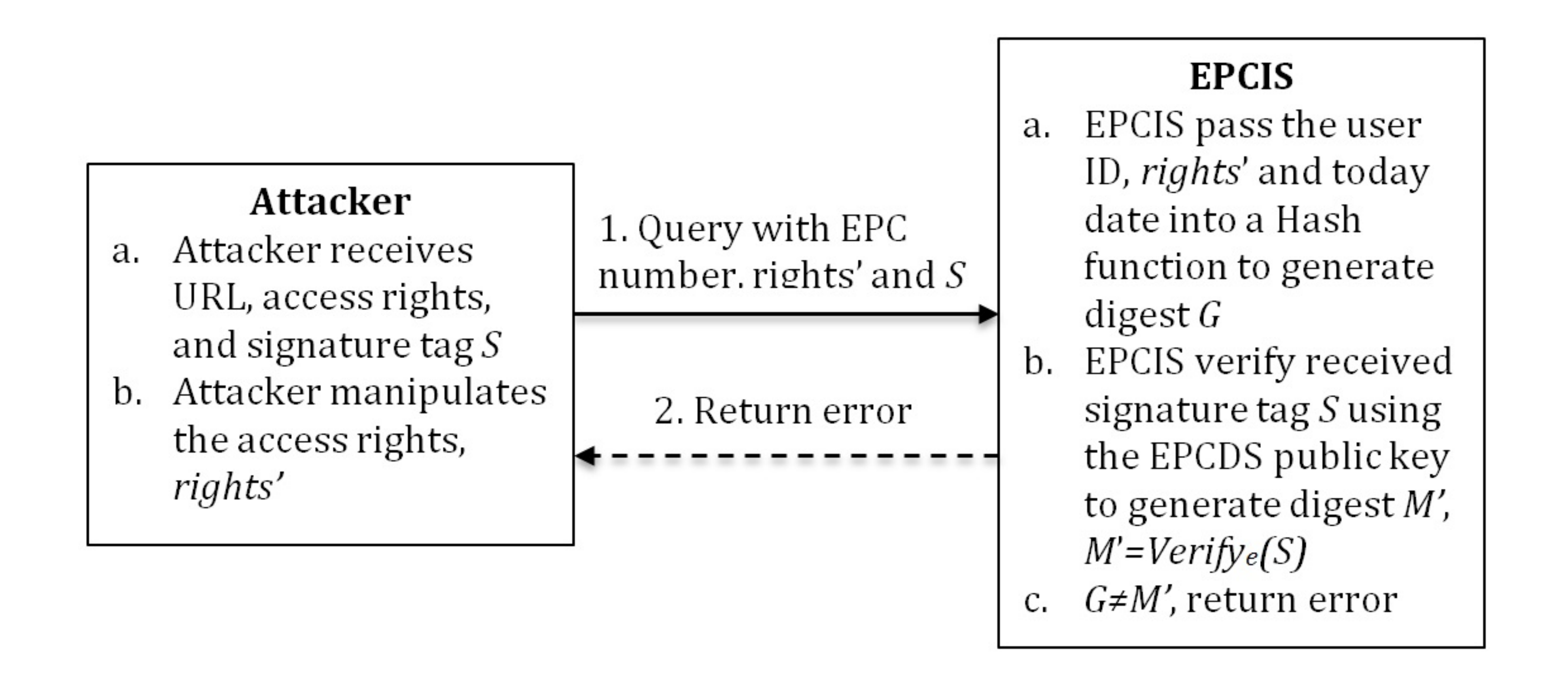}
\caption{Attacker Manipulate Access Rights \label{fig:rights}}
\end{center}
\end{figure}

\textbf{Manipulating Access Rights}. Figure~\ref{fig:rights} shows a simple obvious attack, where an attacker could try to change the access rights received from EPCDS and send it to the EPCIS. In this scenario, the EPCIS will generate a message digest \emph{G} using the attacker's user ID, received access rights and today's date. EPCIS then verifies the received signature tag \emph{S} using EPCDS public key to generate a message digest \emph{M'}, the generated message digest \emph{G} will not match the message digest \emph{M'}, i.e.\emph{G} $\neq$ \emph{M'}. This is because the tampered access rights was not signed by EPCDS. This thus triggers an alarm and the EPCIS will return an error to the attacker. 

\begin{figure}[h]
\begin{center}
\includegraphics[scale = 0.25]{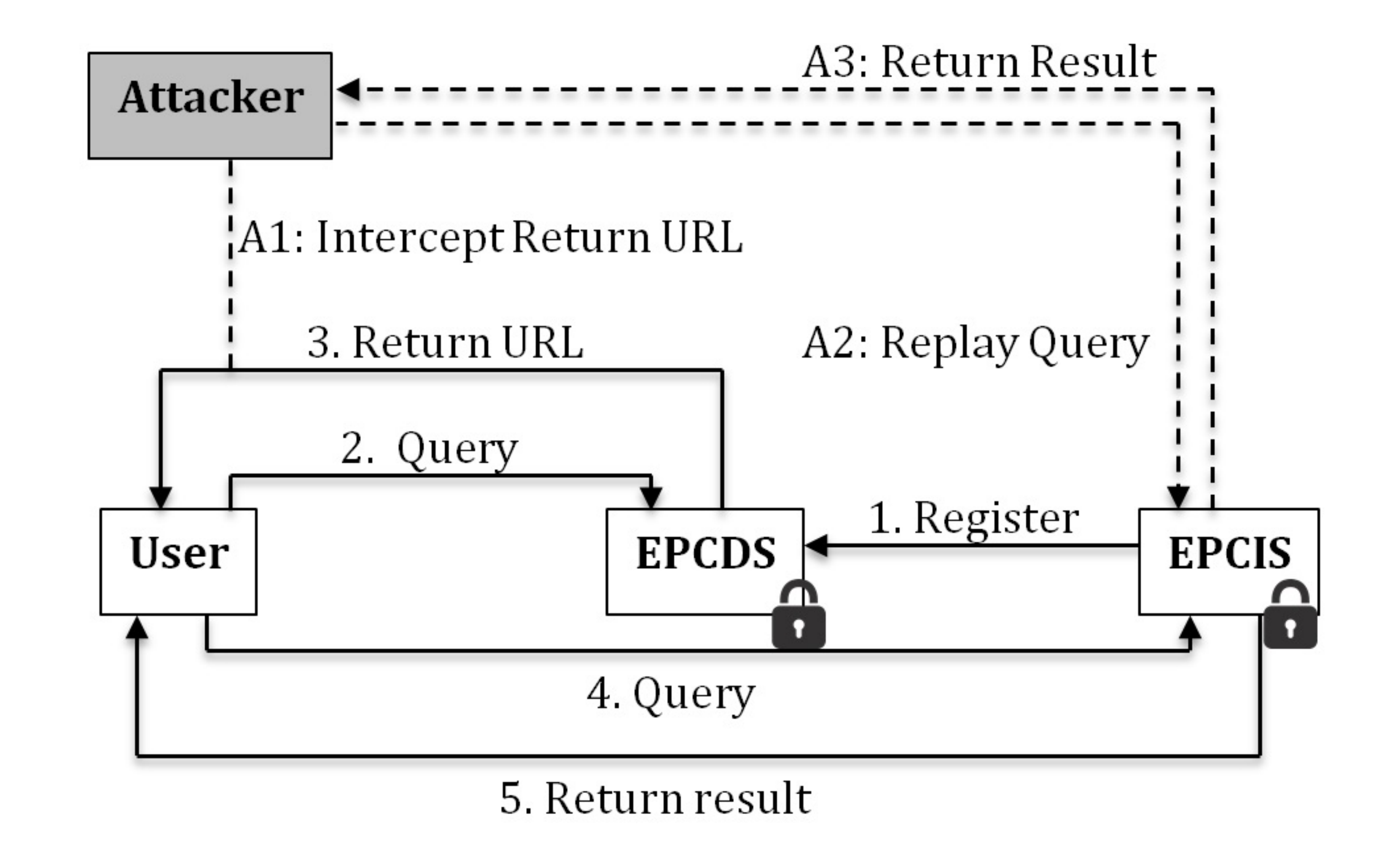}
\caption{Replay Attack on Secure EPCDS Model\label{fig:replay}}
\end{center}
\end{figure}

\textbf{Replay Attack}. The attacker may also attempt more advance attacks such as the replay attack. Figure~\ref{fig:replay} shows the anatomy of a replay attack on the Secure EPCDS Model. A user first query the EPCDS with a certain EPC number and the EPCDS verifies the user against the access control policies and returns the URL to EPCIS if the user has the access right to the EPC event information. An attacker then intercepts the return URL from the EPCDS query of a legitimate user and replay the URL to query EPCIS for EPC event information.  

A simple way to counter the replay attack is to have the EPCIS verify the access control rights of the querying user with EPCDS and this is implement in the Secure EPCDS Model. However in this approach, the access control bottleneck at EPCDS still persist.

The SignEPC model will be able to tackle replay attack using the signature verification mechanism. When an attacker attempts an replay attack on the EPCIS by replaying the query of a legitimate user, he would also have to send the access rights and signature tag of the legitimate user to EPCIS. 

\begin{figure}[h]
\begin{center}
\includegraphics[scale = 0.23]{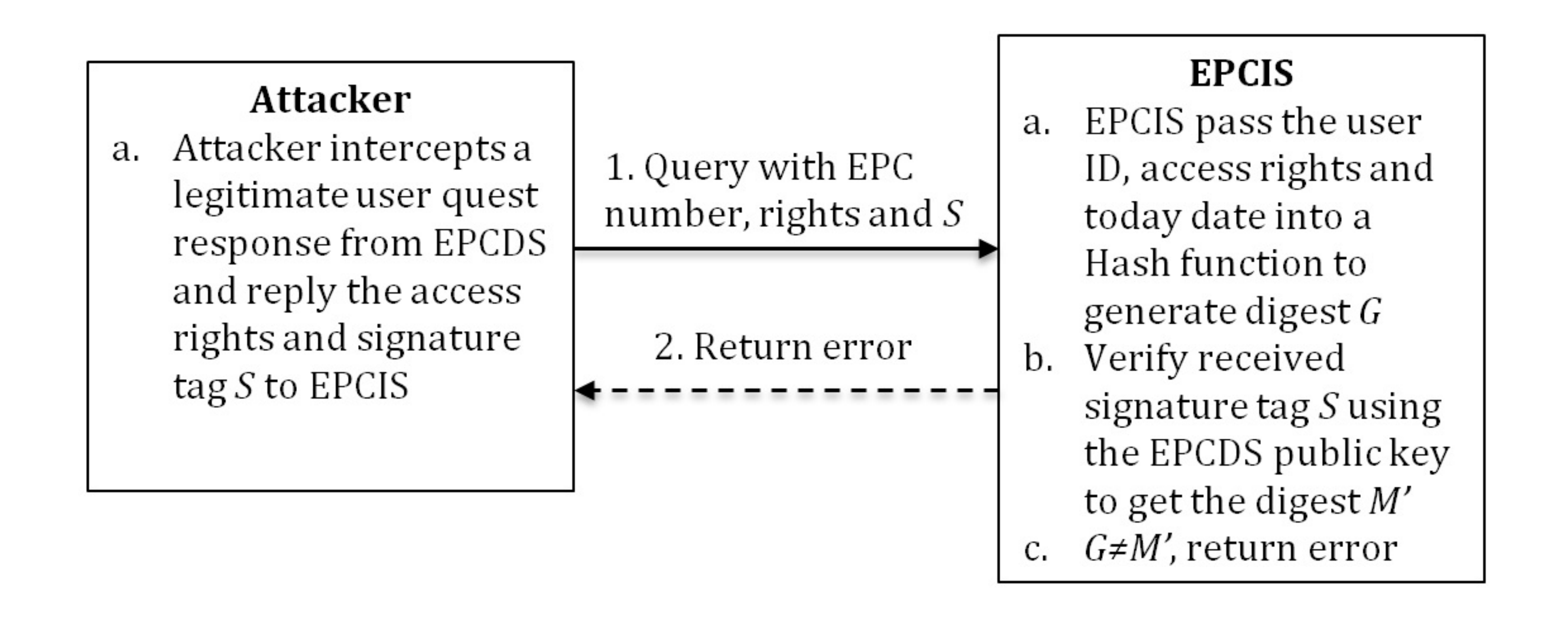}
\caption{Attacker Replay Query to SignEPC \label{fig:tackle}}
\end{center}
\end{figure}

Figure~\ref{fig:tackle} shows the SignEPC signature verification mechanism preventing the replay attack. The EPCIS first pass the user ID, received access rights and today's date into a hash function to generate message digest, \emph{G} = \emph{Hash(userid, rights, todaydate)}. EPCIS then verifies the signature tag \emph{S} to generate a message digest \emph{M'} using the EPCDS public key \emph{e}, i.e. \emph{M'} = \emph{verify\textsubscript{e}(S)}.  In this scenario, the different user ID inputted in the hash function will result in a different message digest, thus \emph{G} will not match the verified message digest \emph{M'}, i.e. \emph{G} $\neq$ \emph{M'}. This will trigger an alarm and return an error to the attacker.

An assumption made in the above scenario is that the attacker is not able to impersonate the legitimate user but only forwards a legitimate query to EPCIS, i.e. the attacker must send his user ID to EPCIS or the user is already authenticated using another authentication mechanism. In business context, we assume that the attacker is a legitimate and authenticated partner but he does not have the access to the queried EPC event information, thus the EPCIS is able to authenticate the attacker even though the attacker does not have the authorization to the EPC event information. 

\subsection {Performance and Scalability}
We analyze performance in two cases - Firstly we compare the time taken by EPCDS to return the URLs to the querying user. Secondly, we compare the time taken by the EPCIS to check if the querying user is authorized to retrieve the EPC event information. 

In the first case, EPCDS in the SignEPC model will take longer time to return the results to the querying user because after validating the user has the access rights, it has to generate a tag that vouch for the user access rights using EPCDS private key. The original Secure EPCDS model will be relatively faster because EPCDS will simply return the URLs to the querying user after the user is verified to have the access rights.

In the second case, EPCIS in the Secure EPCDS Model will have to query EPCDS to check if the user is authorized to retrieve the EPC event information. In the SignEPC model, EPCIS will not need to query the EPCDS again to check for access rights, instead it will verify the access rights of the querying user using the EPCDS public key. Is it not obviously clear which model takes a shorter time to verify the querying user access rights. However as the number of EPCIS increases, a bottleneck is created at EPCDS in the Secure EPCDS Model due to the increase in access control queries. In this scale-up scenario, the Secure EPCDS Model could take longer time in verifying the access rights of the querying user than the SignEPC model. Thus the elimination of the access right bottleneck at EPCDS makes SignEPC more scalable than the Secure EPCDS Model. 

\section{Conclusion and Future Works}
\label{sec:conclusion}
In summary, we have designed and proposed SignPEC, a digital signature scheme for access control solution for EPCglobal network. We have considered the possible attacks that could be done on SignEPC and demonstrate how SignEPC can prevent these attacks effectively. We have also evaluated the performance and scalability of SignEPC, and shown that SignEPC model would outperform Secure EPCDS Model as the number of EPCIS increases. For future works, we will like to explore applying the RSA public key cyptosystem to enhance other security aspects of EPCglobal network. For example we can further improve the security of SignEPC by exploring the use of RSA public key cyptosystem to create a new authentication mechanism that prevent attackers from impersonating legitimate users.

\bibliography{ref} {}
\bibliographystyle{abbrv}

\end{document}